# What is the T-Algorithm? A case study to evaluate a new University

Jose Berengueres
College of IT
UAE Univerisity
17551 Al Ain, A.D, U.A.E
jose@uaeu.ac.ae

*Abstract*— We evaluate Scott Galloway's T-algorithm as a thinking framework via a simulated case. We explain how we applied it to analyze the investment strategies when starting a new university. We note that the algorithm can be used to not only describe but also prescribe the strategy design space. We also point out the weak and strong points of the said algorithm and compare it to existing tools such as the Business canvas model and SWOT.

## I. Background

In this section, we provide **key** background information that we will apply later to illustrate the application of the T-algorithm with a simulated case where the user has to decide the strategy for a new university. You may skip this section if you are familiar with the educational sector. The T-algorithm is a thinking framework by Scott Galloway. In this paper, we will evaluate it from a visualization point of view.

### A. The three types of Educational Institutions

In the past three decades (1989-2020), many government and private initiatives have started **new** universities. Yet, few have succeeded in becoming a resounding success like Stanford. For a summary of the serendipitous ways of *"How Stanford became Stanford"* see the talk by Steve Blank [10]. We can classify these new universities into 3 groups by type.

**Campus-based** - Examples are the Singapore University of Design and Technology (SUDT), the Dubai Institute of Design and Innovation (DIDI), and the Hasso Plattner Institute (Potsdam). These new universities have had different approaches to education yet all of them have had one strategy in common, they have invested in **physical campuses** and face-to-face lecture-style teaching with full-time faculty. On the other hand, we have a new breed of "universities" that are digital-first.

**Digital-first** - These are Udacity, Udemy, Coursera, and Lynda.com (now LinkedIn Ed.). These digital universities are focused on teaching everything related to digital media arts and neighboring subjects. Additional examples of spinoffs (skunkworks) from established institutions are Harvard's EdX, Hasso Plattner Institute online, and so on. These two later ones are by institutions that have gone online while at the same time developing their own proprietary online learning platforms. Thus, also entering into direct competition with the earlier platform providers such as Pearson's Blackboard, the open-source Moodle, and the Coursera, Udacity, and Lynda platforms.

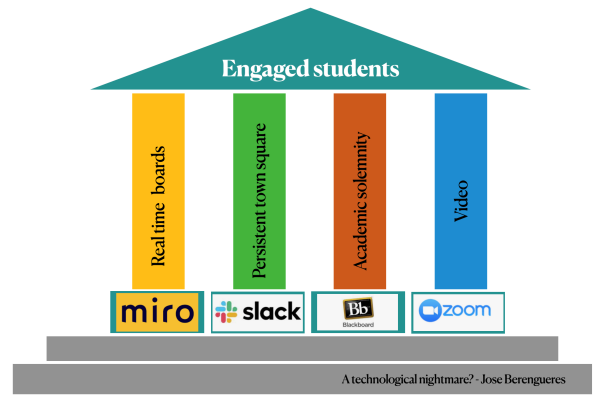

*Fig. 1 The four pillars of online education. Source: Author.*

Then we have the niche players /outliers. Note that these platforms provide only one of the four elements required for efficient online education (video conferencing, persistent town square, solemnity in certification, real-time whiteboarding) See Fig. 1

**Niche players** – These are small-sized academies that focus on a niche area and excel at teaching it. Examples are the Copenhagen Institute of Interactive Design (CIID), Harbour. Space in Barcelona, Kyiv Academy of Media Arts (KAM), and Stockholm's Hyper Island (HI). HI is the pioneer in the segment; founded in 1996 [1]. Both HI and KAM were started by **Admen** and are more vertically integrated than the others because the close relationship with the industry makes them efficient at job placing. One differentiator of HI and KAM vs. the rest is the final project style, which combines

(i) an internship,

(ii) recruiting and job placing and,

(iii) a final project in one step.

HI has also no full-time faculty or teachers [1]. Finally, there is a fourth class. Institutions from group 1 that have reinvented themselves. The hallmark example, in this case, is RISD's transformation from analog to digital arts under John Maeda, whose leadership propelled RISD to the top 5 worldwide in its category. In this case, the word digital hardly appeared in the syllabi of RISD pre-Maeda and it was found often after he took leadership [2]. Finally, we note that in the VC arena there are examples of vertical integration of schools and funding. An example is "Antler.co". Antler is an accelerator that offers funding that is tied to compulsory training provided and supervised by Antler. Antler is also known as the YC of EU [11] and has had a strong process-oriented culture that was inherited from its McKinsey alumni (3 of the 4 co-founders are former McKinsey). In summary, new universities that are campus-based have not been particularly successful while digital-first operations are mainly profitable. For example, it is estimated that Coursera made 140 Million in revenue in 2018 [3]. Niche players continue to be niche players. With the exceptional profitability of KAM in Ukraine; and the growth progression of HI, that now has expanded to various continents without major media coverage.

*B. The study case*

Let us assume we are a city-state government with the mandate to start a new university or revamp our own educational system, how should we do it? We use Scott's Galloway T-algorithm to visualize, clarify and evaluate. After a first iteration we assume that our new university would like to enable the following success factors found in the algorithm (See table 1):

- Be part of a **Rundle (recurring revenue bundle)**, either from the government in form of certification / civil service employment regulation and or favoritism or via a long life learning scheme (part of public policy) or via private industry close partnerships that make employability of graduates smoother.
- Be vertically **integrated**, i.e. To be 100% effective at placing graduates in a job. Note that most MBA's do not accomplish this **critical** task for 10 to 30% of the cohort after one year of graduation.
- **Scale with data**. Becomes better with Data. This means the leadership of the university might be someone who stewards a data culture. For example, Stanford professors co-founded Coursera.

## II. T-ALGO

In this section, we apply the said algorithm proposed first by Scott Galloway circa 2019. The said algorithm is usually found as two Excel tables that recommend an investment strategy regarding 8 KPIs. These KPIs are the following, see also Table 2:

1. Does your product appeal to human instincts and passions (procreation, security, belongingness, or basic needs?)
2. Is the company a top place to start a career (is it a McKinsey? If not, how to attract top talent?)
3. Has the company had healthy growth and profit margins?
4. Has the product potential to be wanted as part of a bundle (example: would Netflix be interested in buying your movie for their subscription?)
5. Are you leveraging Vertical integration?
6. Can it leverage Software economies of scale?
7. Does it have a compelling storytelling that inspires?
8. Is the leadership liked or disliked on social media?

Table 1 shows the first of such tables, called the ***investment grid***. It compares the considered company (in our case a new University) to peers and it is based on qualitative estimations by the user. The peers or competitors are placed in descending ordinal order for each of the 8 given KPIs. Table 2, is called the ***differentiation grid***. This grid helps us summarize the data in Table 1 by reducing information overload and recommending where to invest.

*A. What does it recommend?*

For each KPI the algorithm recommends one of two strategies *Tables Stakes* or *Differentiator*. Table stakes (a poker term) means change nothing. The Excel formula used is:

```
IF(COUNTA(J$17:J$22)>COUNTA(J$30:J$32),
   "Table Stakes",
   IF(COUNTA(J$17:J$22)=0,
       "NA",
       "Differentiator"))
```

where the J column is a column of a given KPI of Table 1, (See Table 1). This means that, given a KPI, if there are more competitors in the top 6 cells (top 54%) than in the bottom 3 cells, (bottom 27%), of the given column then the algorithm recommends **Table stakes**, otherwise, it recommends using the KPI as a **Differentiator** in the investing strategy (capital allocation). In addition, Table 2 also classifies the proposed company (in our case 'a new University'), into three classes depending on the following code also found in Table 2.

```
=INDEX($A$15:$A$32,
   MATCH('New University',
   C$15:C$32,0))

   Where ($A$15:$A$32) is an array:
   {"Advanced Core Competence"
   "Intermediate, progress made, needs improvement"
   "Novice/Not on Radar"}
```

This Excel code means that if your company (in our case a new University), is placed in the top 6 in a given KPI column Table 1, then it is mapped to the label ***advanced***, if placed with the bottom 30%, it is mapped to the label ***novice*** and otherwise defaults to ***intermediate***. The rationale for these rules is not clear. Perhaps they can be justified by a winner takes all thinking framework. However, this rule is of course, not general as the history of hit products has something else to say (See the history of Apple, Amazon, Google, f.r.o.g design, Toyota, Sony Walkman, Sony PlayStation, Tesla, Google, and so on); in addition, many products that followed this rule have failed (See General Electric, DeLorean cars or Cannondale bicycles for instance). However, a tool is as good as its input data, and the real value of the tool is perhaps to visualize and be aware of the relation between product leadership and the 8 KPI areas.

*B. Note on the naming*

Regarding the naming of KPIs. These KPIs are not novel. For example, the 'appeal to instincts' is the equivalent of the

Silicon Valley adage "sell me painkillers, not vitamins" meaning the benefits of selling products in markets with inelastic demand (pharma, education, transportation, real estate, search, groceries & food). The 'Benjamin Button effect' that refers to products that improve with time and user numbers is the old network effects or what a16z calls 'software is eating the world'. The vertical integration advantage was summarized by Steve Jobs a decade ago when he realized that Apple needed to own and produce its silicon chips to not depend on Intel or Samsung. It is summarized in his quote "Own your Silicon". And so on.

Table II 8 KPIs for a New University as envisioned by the author.

| Kpi | Our New University | Investment level recommended by T-algo |
|---|---|---|
| 1. Appealing to Human Instinct |  | Table Stakes |
| 2. Accelerant | Novice/Not on Radar | Table Stakes |
| 3. Maintaining Growth + Margins | Novice/Not on Radar | Table Stakes |
| 4. Rundle | Novice/Not on Radar | Table Stakes |
| 5. Vertical Integration | Advanced | Differentiator |
| 6. Benjamin Button (scaling via software) | Advanced | Differentiator |
| 7. Visionary Storytelling | Advanced | Differentiator |
| 8. Likeability | Intermediate | Table Stakes |

Labels that the t-algo decides based on data from Table 1

### III. Discussion

The t-algo, like any other visualization framework or thinking tool [6] has helped visualize, clarify, and storytelling strategy. It is, perhaps, a modern successor of the traditional SWOT chart with some rules attached to it. However, the large number of categories and scattering in two tables overflow the *7-chunk memory limit* of humans, and thus it might benefit from a simplification if it wants to become as popular as Osterwalder's biz canvas [5] (9 categories in one A4 or the A3/PDCA, one report one A3 size sheet of paper). Note how other famous charts that changed the world [6-8] never violate the 7-chunk rule. See Fig. 3.

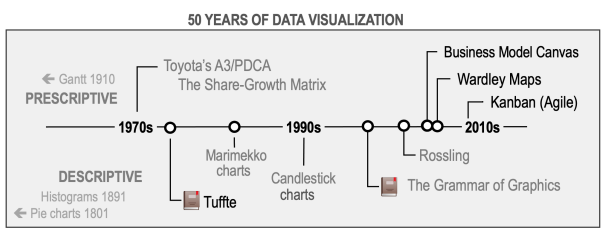

*Fig. 2 What is the common factor in visual thinking methods? Source: [6]*

Finally, we note that the t-algo is useful not only to **ideate** but also to iterate the **concepting** of a strategy or product. In this case, for example, we note that our new university has a lot of unexploited potential in likeability; and we note that other competitors like harbour.space have leaders with high likeability (Svetlana Veliakonova), with underused potential. Next, we list the insights gained from using the t-algo so the reader can appreciate the value of it. This took about 2 hours of work, an amount of time comparable to the same exercise with the business canvas model.

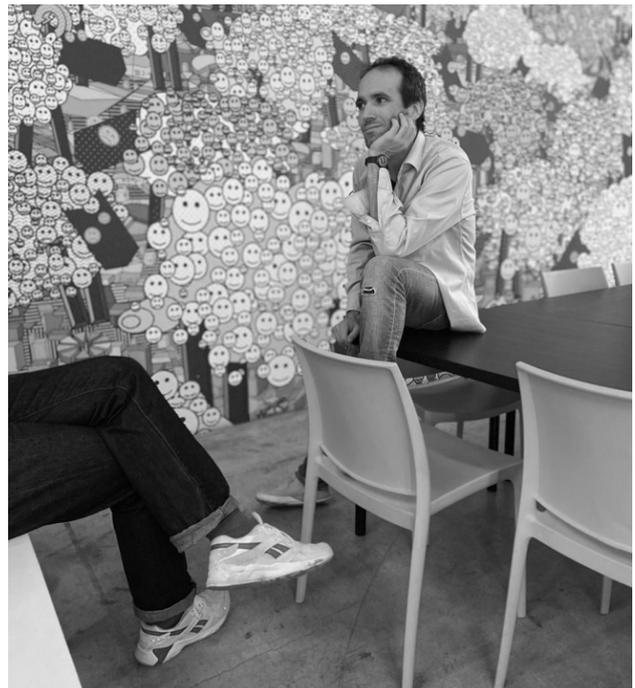

*Fig. 3 Example of use of art to affect likeability. KAM campus has invested in funky art like this wall made entirely with Excel. Photo: Yulia K.*

#### A. Main takeaways

*1) Likeability*

We note that *likeability* can be not only about the leader but also about the institution. In the case of KAM, for instance, we have given high likeability score due to its campus which in our opinion has surpassed the gold standard in the interior design of HI and also contributes importantly to the likeability of the institution (See Fig. 2). Note that this idea is not new. A few traditional MBA schools such as Aalto Executive in Helsinki and even hospitals such as Mayo Clinic in Florida are leaders in using art on campus to improve perceived likeability.

Another insight from filling the tables is that new universities that are free from the publishing pressure of academia can focus on hiring more likeable faculty in ways traditional universities cannot. Thus, a competitive advantage of the past (high impact paper producing faculty) has become a comparative hurdle as the student experience and employability impact more and more the (online) reputation of the university, and the benefits, drawbacks, and ROI of an MBA become more transparent [8, 9].

*2) Rundle & Vertical integration.*

Currently, no universities offer a rundle. Reputable MBA schools offer some sort of rundle in combining the admission to the program with an automatic loan to pay for the MBA fees and a reasonable promise of a high-paying salary or salary bump upon completion.

*3) Benjamin Button*

From direct experience, we can say that no university today is large enough to qualify for big data. On the other hand, very few universities, except perhaps Coursera and similar, use analytics as productively as Google or Spotify do.

*4) Storytelling*

A new university, because it is more agile and is not subject to old legacy investments and sunk costs, has the capability to focus on areas of growth in the new economy in ways traditional universities cannot. As we saw from the introduction, all the new universities that have gone the traditional route have not been a resounding success while only niche players and digital-first ones have had significant success. This, obvious in retrospect, is summarized in the adage: run towards where the ball is going to be, not towards where it is now.

## IV. CONCLUSION

The T-algo can be applied for descriptive as well as prescriptive analytics and concepting in a way similar to the Business canvas model and the SWOT chart. The T-algo's KPIs can be expanded to broader areas. For example, likeability need not only to be circumscribed to the CEO's online reputation but can also encompass the interior design and design language of the organization. See also gaining the goodwill of the community in [4]. Overall, the T-algo is useful in detecting strategy **blind spots**. That is not a strong point or the purpose of the Business canvas model or SWOT. While the Business canvas model excels at innovation by "component shifting", the T-algo seems more adept at innovation by paying attention to the competition. However, 2D thinking frameworks such as Gartner's magic quadrant are powerful contenders when it comes to the strategy design space and competitor analysis [12,13]. Ironically, this kind of competitor analysis is something that would go against Jeff Bezos's success story and philosophy: "*obsess about the customers, not the competition*".

ACKNOWLEDGMENT

With input from Gabe Fender, Bianca Morris, James Lackland, Moana Tumahai, Nadav Gur, and Daniel Plant.

Table I. T-Algo Grid

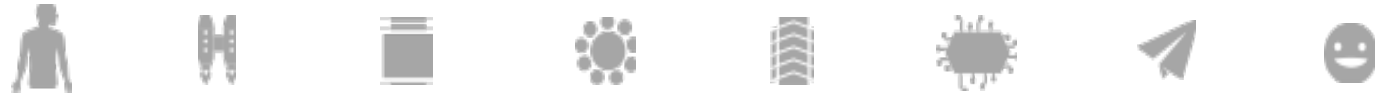

| | Appeals to Human Instinct | Career Accelerant | Growth + Margins | Rundle | Vertical Integration | Benjamin Button Product | Visionary Storytelling | Likeability |
|---|---|---|---|---|---|---|---|---|
| **Advanced Core Competence** | EdX / Harbour.Space / HI | KAM / HI | KAM / HI | My New Uni | KAM / HI / My New Uni | EdX / Udemy / Coursera / My New Uni | EdX / HI / Harbour.Space | Harbour.Space / KAM / HI |
| **Intermediate Concrete progress made, but needs improvement** | KAM / CIID | Harbour.Space / CIID | EdX / Udemy / Coursera | | | HI | KAM / My New Uni / CIID | My New Uni / CIID / EdX |
| **Novice/Not on Radar Experimenting with, Bad at, OR Not on radar** | Udemy / My New Uni | EdX / Udemy / My New Uni | CIID / Harbour.Space / My New Uni | CIID / KAM / HI / EdX / Udemy / Harbour.Space | CIID / EdX / Udemy / Harbour.Space | CIID / KAM / Harbour.Space | Udemy | Udemy |

Table III. – T-Algo investment grid – how to spend it.

| | Not Applicable | Differentiator | Table Stakes |
|---|---|---|---|
| **Advanced** Core Competence | | Maintain investment as you've identified a way to differentiate from competitors. Continue to monitor ROI | Maintain investment. Don't stop innovating or accelerating here as competitors will continue to invest. |
| | | **Vertical integration , Rundle, BB** | **Visionary St, Growth+margins** |
| **Intermediate** Concrete progress made, but needs improvement | Decrease investment unless you think this is truly a differentiator | Maintain investment as you're investing in a potential differentiator. Continue to test & learn and decide to accelerate/decelerate. | Maintain investment, and increase investment if resources allow |
| | **Accelerant, growth, appeal to hi** | | |
| **Novice/Not on Radar** Experimenting with, Bad at, OR Not on radar | Maintain investment, don't invest in pillars that aren't applicable to your company | Increase investment if all table stake strategies are a core competence or you have no differentiators. | Increase investment. Put this strategy on your roadmap and implement immediate next steps to start making progress. |